# Ultralow-frequency collective compression mode and strong interlayer coupling in multilayer black phosphorus

Shan Dong,[1] Anmin Zhang,[1] Kai Liu,[1] Jianting Ji,[1] Y. G. Ye,[2] X. G. Luo[2,6] and X. H. Chen[2,5,6]

Xiaoli Ma,[1] Yinghao Jie,[1] Changfeng Chen,[3] Xiaoqun Wang[1,4] and Qingming Zhang[1,4*]

[1]Department of Physics, Beijing Key Laboratory of Opto-electronic Functional Materials & Micro-nano Devices, Renmin University of China, Beijing 100872, P. R. China

[2]Hefei National Laboratory for Physical Sciences at Microscale and Department of Physics, University of Science and Technology of China and Key Laboratory of Strongly-coupled Quantum Matter Physics, Chinese Academy of Sciences, Hefei, Anhui 230026, China

[3]Department of Physics and High Pressure Science and Engineering Center, University of Nevada, Las Vegas, Nevada 89154, USA

[4]Department of Physics and Astronomy, Collaborative Innovation Center of Advanced Microstructures, Shanghai Jiao Tong University, Shanghai 200240, P. R. China

[5]High Magnetic Field Laboratory, Chinese Academy of Sciences, Hefei, Anhui 230031, China

[6]Collaborative Innovation Center of Advanced Microstructures, Nanjing 210093, China

**The recent renaissance of black phosphorus (BP) as a two-dimensional (2D) layered material has generated tremendous interest in its tunable electronic band gap and highly anisotropic transport properties that offer new opportunities for device applications. Many of these outstanding properties are attributed to BP's unique structural characters that still need elucidation. Here we show Raman measurements that reveal an ultralow-frequency collective compression mode (CCM), which is unprecedented among similar 2D layered materials. This novel CCM indicates an unusually strong interlayer coupling in BP, which is quantitatively supported by a phonon frequency analysis and first-principles calculations. Moreover, the CCM and another branch of low-frequency Raman modes shift sensitively with changing number of layers, allowing an accurate determination of the thickness up to tens of atomic layers, which is considerably higher than those previously achieved by using high-frequency Raman modes. These results offer fundamental insights and practical tools for exploring multilayer BP in new device applications.**

The successful mechanical exfoliation of few-layer graphene from bulk graphite marks the beginning of a new era in 2D layered materials and their applications in nanoscale devices. Few-layer graphene offers many exciting properties such as high mobility, low resistivity and high thermal conductivity, as well as excellent mechanical performance, which are suitable for countless potential applications in ultrathin optical, electronic and mechanical devices[1]. The zero band gap of graphene, however, greatly limits its applications in logic devices. This has stimulated a concerted effort to search for alternative 2D materials, such as transition-metal dichalcogenides (TMDs), which exhibit intrinsic finite band gaps; but TMDs reported so far only show mobility up to a few hundred $V^{-1}s^{-1}$ that is insufficient for practical applications[2].

The rediscovery of BP from the perspective of a new 2D layered semiconductor has cast promising light on offering a new range of key physical properties, such as the band gap and mobility, needed for applications. Similar to graphene, BP has a corrugated honeycomb lattice[3] with strong intralayer covalent bonding and weak interlayer van der Waals coupling[4]. It has an orthorhombic crystal symmetry $D_{2h}$ with a layer-layer distance of ~5.3 Å[5], as illustrate in Fig. 1b. BP also possesses widely tunable band-gaps of 0.33-2.0 eV from bulk to few-layer forms[6]. Its room-temperature mobility can reach up to ~1000 $V^{-1}s^{-1}$ in exfoliated flakes with a thickness of ~10 nm[7]. In addition, BP exhibits a high on-off ratio and good mechanical properties. These properties are suitable for electronic and opto-electronic applications[8,9].

Raman scattering is a versatile technique for investigating fundamental lattice dynamics, electronic and excitonic properties of low-dimensional materials. For layered compounds, there exist two basic modes reflecting relative motions between neighboring layers, i.e., the compression/breathing and shear modes. These modes offer essential information about the interlayer coupling, which plays important roles in probing the lattice dynamics and electronic properties of few-layer graphene and $MoS_2$[10-12]. The Raman frequencies shift with the flake thickness, thus allowing an accurate determination of the layer number; such shifts also measure the interlayer electron hopping crucial to heterojunction- and quantum-dot-based applications[13,14].

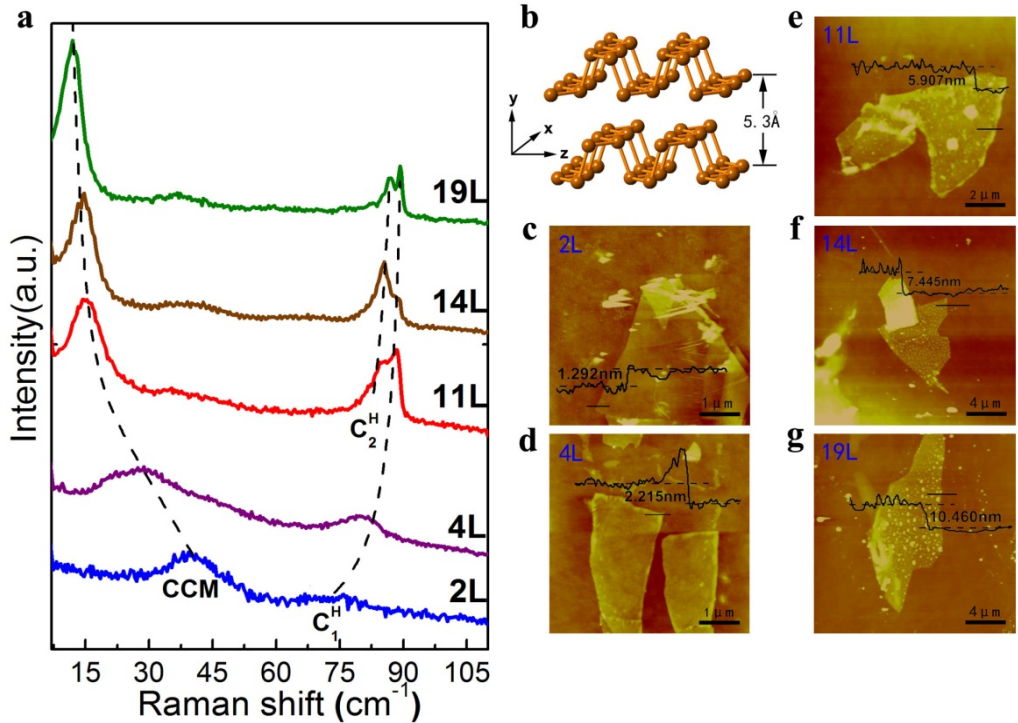

**Figure 1 Raman spectra of mechanically exfoliated few-layer BPs.** (a) Two branches of Raman modes at low (~100 cm$^{-1}$) and ultralow (~10 cm$^{-1}$) frequencies with changing layer number. (b) an illustration of the layered BP structure. (c)-(g) The layer thickness of the selected flakes is accurately characterized by AFM imaging in each case.

In this paper, we report observation of novel Raman modes with very low frequencies in multilayer BP. There are two distinct branches of such modes, both identified as compression modes according to symmetry analysis, one with low frequencies that approach 100 cm$^{-1}$ with increasing layer number and the other with ultralow frequencies that approach 10 cm$^{-1}$, which is unprecedented in layered 2D materials such as graphene and TMDs. While the low-frequency (LF) mode exhibits a clear layer-number (N) dependence described by a linear-chain model, the ultralow-frequency (ULF) mode scales with the layer number in the limit of large interlayer coupling, suggesting an unusually stronger interlayer coupling in multilayer BP. Our phonon frequency analysis and first-principles calculations provide further evidence of strong interlayer coupling in BP. The present approach establishes a sensitive and robust method for an accurate determination of layer thickness of atomically thin BP films, and the findings about the unexpectedly strong interlayer coupling in BP imposes important constraints on the modeling and design of BP-based device applications.

To establish an accurate relation between the layer number and phonon modes, five BP flakes with different thicknesses are selected for Raman measurements and simultaneous atom force microscopy (AFM) imaging (Fig. 1), and the AFM directly gives the thickness of each flake. In the Raman spectra for each flake, two phonon branches, namely the LF and ULF Raman modes, are observed. These two branches of Raman modes evolve differently with increasing layer number: not only they shift in opposite directions as the layer number increases with the LF mode moving to higher frequencies while the ULF mode moving to lower frequencies, but more importantly they follow qualitatively different scaling behavior (see below for a detailed analysis). Meanwhile, the widths of these modes consistently become narrower with increasing layer number.

The significant changes in both the frequency and width of the LF and ULF Raman modes can serve as a clear and accurate indicator of the layer number in the BP flakes. To quantify this observation, we have fit the layer-number dependence of the mode frequencies according to a linear-chain model (see details below). For a high-quality fitting, we have made additional Raman measurements on a large amount of BP flakes as shown in Fig. 2(a). The smooth evolutions of both the frequencies and widths of the observed modes allow accurate assignments for all the obtained spectra. Their frequencies follow two well-established fitting curves, thus allowing an accurate determination of the layer number [see Fig. 2(b) and (c)]. We further checked the layer-number dependence of the high-frequency (~470 cm$^{-1}$) $A_g^2$ mode and obtained consistent results (Supplementary Information). It should be noted that the maximum shift in frequency for the $A_g^2$ mode is limited to ~3 cm$^{-1}$ and the mode frequency nearly stops changing in samples with more than five layers[15]. By comparison, the maximum frequency shift in our measurements is as large as 32 cm$^{-1}$ for the ULF mode and 19 cm$^{-1}$ for the LF mode, and the mode positions are still sensitive to the thickness change even beyond several tens of layers. This observation demonstrates that the ULF and LF Raman modes in BP offer a highly effective method to determine the thickness of atomically thin films over a wide range.

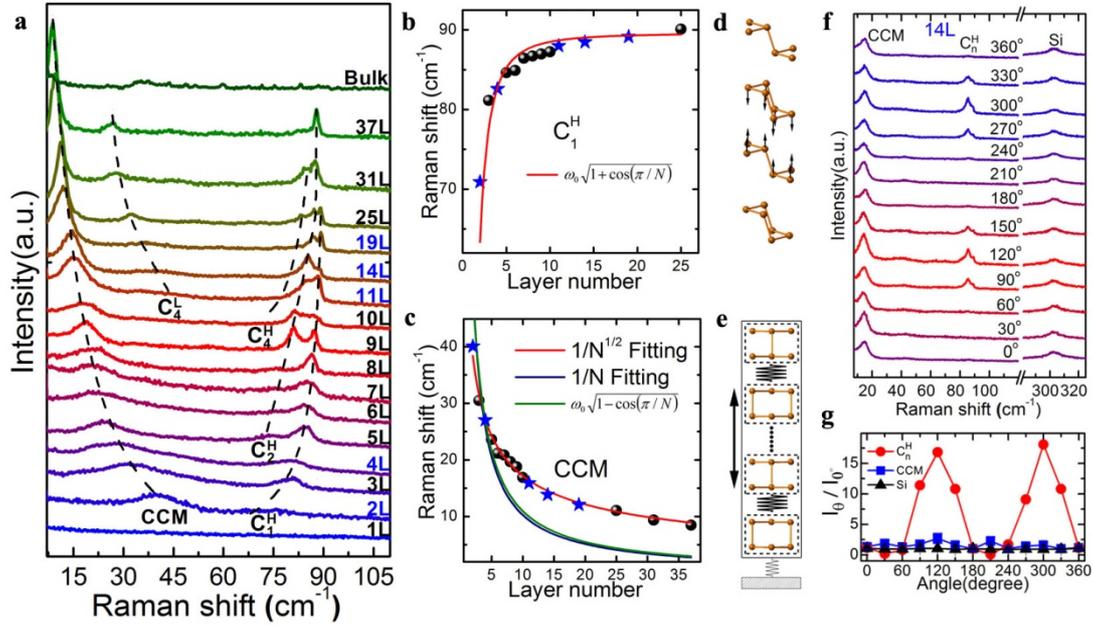

**Figure 2 The LF and ULF modes and the linear-chain fitting.** (a) Raman spectra collected from few-layer to bulk BPs. The blue layer number means that the corresponding flake thickness is checked by AFM imaging. CCM, $C_n^H/C_n^L$ denote the collective compression mode, the nth-order compression modes from the higher (LF)/lower branch, respectively (See the text and Fig. 4). (b) Layer-number dependence of the $C_1^H$ mode. The corresponding vibration pattern is illustrated in (d). The data are extracted from the Raman spectra in (a), and those represented by solid stars from those marked by blue layer numbers, which are confirmed by AFM imaging. The red solid curve is the linear-chain fitting. (c) Layer-number dependence of the CCM mode. The corresponding vibration pattern is illustrated in (e). The data are obtained in the same way as in (b). The red, blue and green solid curves are the fittings by $\frac{1}{\sqrt{N}}$, 1/N and the standard linear-chain model. (f) Polarization measurements of the Raman modes. Both the intensities and frequencies are carefully monitored and calibrated by the second-order silicon mode at ~302 cm$^{-1}$. The angle dependence of the integrated intensities of the observed modes is shown in (g).

The LF and ULF modes, which are absent in both bulk and monolayer cases, are assigned to interlayer modes, as previously observed in h-BN, few-layer graphene and MoS$_2$[10-12]. Symmetry analysis further indicates that unlike graphene and MoS$_2$, the interlayer shear modes (B$_{1g}$ and B$_{3g}$) in few-layer BPs are prohibited under our backscattering configuration[5,16,18]. First-principles calculations and symmetry analysis demonstrate that both the LF and ULF branches observed in BP come from interlayer compression motions (Supplementary Information).

As shown in Fig. 2(b), the layer-number dependence of the mode frequencies of the LF branch follows the prediction of a standard linear-chain model, which gives $\omega = \omega_0\sqrt{1 + cos\,(\pi/N)}$, where ω is phonon frequency, N layer number and $\omega_0$ a fitting parameter. The corresponding atomic vibration pattern is illustrated in Fig. 2(d). The linear-chain model gives[10] $\omega_0 = \frac{1}{\sqrt{2}\pi c}\sqrt{\frac{\alpha}{\mu}}$, where $\omega_0$=71 cm$^{-1}$ is the frequency of the first-order interlayer compression mode $C_1^H$ [Fig. 2(a)], $c$ the speed of light, $\mu$ =1.42×10$^{-26}$ kg·Å the mass per unit cell area. These results lead to an interlayer force constant $\alpha$ = 1.27×10$^{20}$ N/m$^3$, which is in good agreement with a recently calculated value[17]. It is noted that this force constant is significantly larger than its counterpart for MoS$_2$ and graphene[10,11], which suggests a much stronger interlayer coupling in multilayer BP (see Table I). This is completely consistent with the results from the ULF branch (see below). Following $C_{33} = \alpha \cdot t$, where t is the distance between neighboring layers, we obtain the stretching modulus $C_{33}$ ~ 67.3 GPa, which is very close to the value 70.0 GPa measured by neutron scattering and the calculated value 70.8 GPa[18,19].

Surprisingly, the layer-number dependence of the ULF branch shown in Fig. 2(c) clearly does not follow the derived relation $\omega = \omega_0\sqrt{1 - cos\,(\pi/N)}$ or *1/N* from the standard linear-chain model. Instead, it scales as $\frac{1}{\sqrt{N}}$. This scaling behavior is still compatible with the linear-chain model, but only in the limit of large interlayer coupling. In other words, this unusual ULF mode indicates that the interlayer coupling is strong enough to couple all the layers together, resulting in an in-phase compression motion of all the layers in the sample relative to the substrate[20] [Fig. 2(e)]. It can be simply regarded as the collective motion of a composite object having N times the mass per unit area of one BP layer. Such a similar mode has been observed in epitaxial KBr film on NaCl substrate[21]. The assignment of the collective compression mode is also supported by polarization measurements. The LF branch modes exhibit a high degree of anisotropy in intensity, which is in agreement with the observations for the high-frequency Raman modes and photoluminescence[22-24], while the CCM intensity has a much weaker angle-dependence [Fig. 2(f) and (g)], which is consistent with that the CCM is essentially an in-phase vibrational mode smeared by the substrate.

Using the general layer-number dependence given by the linear-chain model, $\omega_N = \omega_0\sqrt{1 \pm \cos{(n\pi/N)}}$, we can make a comprehensive assignment for the observed compression modes except for the CCM mode. These assigned modes follow the fan diagrams (Fig. 4), similar to the case in $MoS_2$[12].

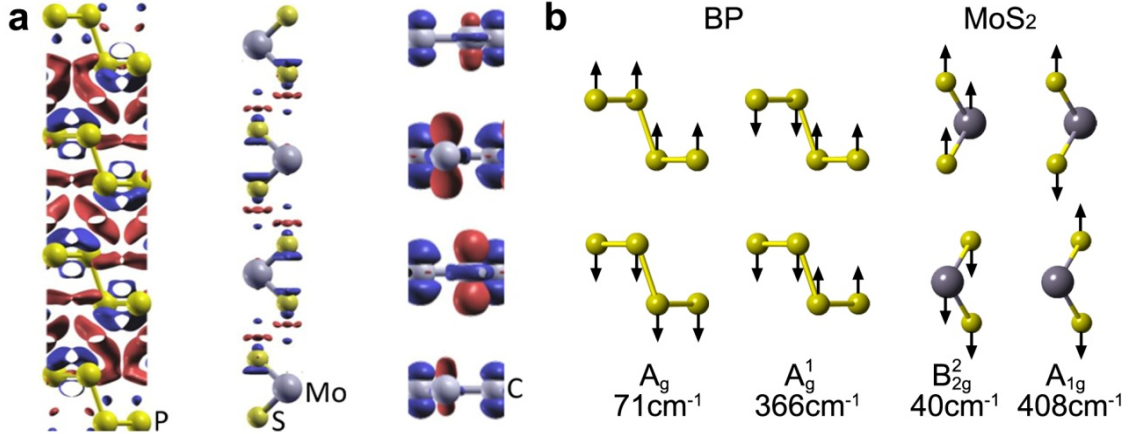

**Figure 3 Charge difference density and phonon mode analysis.** (a) Calculated charge difference density profiles for few-layer BP, $MoS_2$, and graphene. The regions gaining/losing electrons are marked in purple/blue. (b) Displacement patterns and measured frequencies of comparable intra- and inter-layer compression modes in BP (left) and $MoS_2$ (right).

The strong interlayer coupling revealed by the $\frac{1}{\sqrt{N}}$ scaling behavior can be quantitatively estimated by a phonon frequency analysis. The frequency ratio of inter-and intra-layer modes with comparable vibration patterns offers a good measure of relative interlayer coupling, since $\omega^2 \propto K/m$ in a simplified spring model, where K is the spring force constant. Based on the data for the intra- and inter-layer shear modes, the intra-layer spring constant is estimated to be 100 times larger than the interlayer one in $MoS_2$[25]. Using the intra- ($B_{2g}^2$~40 cm$^{-1}$) and inter-layer ($A_{1g}$ ~ 408 cm$^{-1}$) compression modes in $MoS_2$ [Fig. 3(b)], a similar ratio was obtained[11-12]. Following this established procedure, we have identified [Fig. 3 (b)] the $C_1^H$ (71 cm$^{-1}$) mode as the corresponding interlayer version of the high-frequency intra-layer $A_g^1$ mode (366 cm$^{-1}$). We therefore obtain a ratio of ~26 between the intra- and inter-layer spring force constants in BP, which is much smaller than that in $MoS_2$.

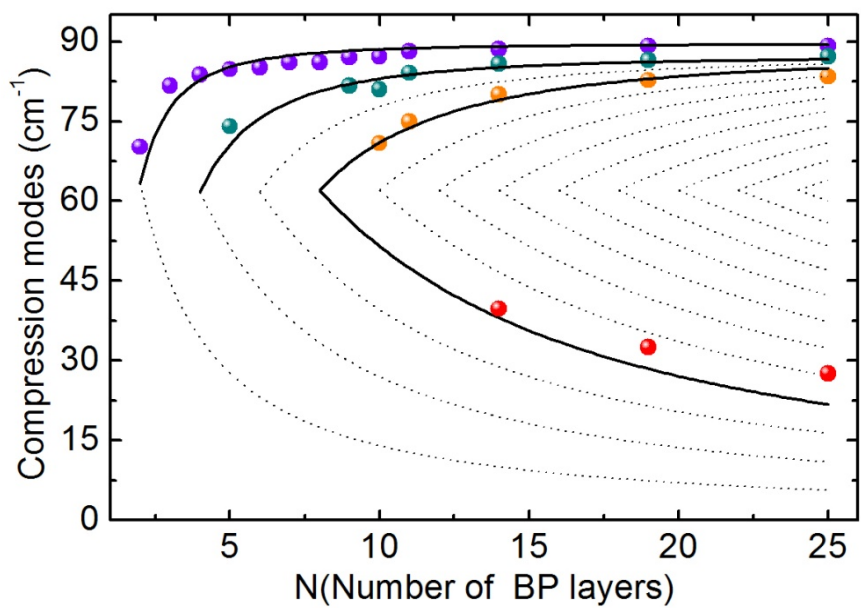

**Figure 4 Assignment of the observed compression modes by fan diagram.** The solid and dashed lines are produced by the linear-chain model (See the text for details). The colored circles are the positions of the observed compression modes in Fig. 2 (a) except the CCM.

We further made first-principles calculations to examine the interlayer coupling. In Fig. 3 (a), the charge difference density profiles are shown for few-layer BP, $MoS_2$, and graphene. More significant charge redistributions can be clearly seen in few-layer BP, which indicates stronger overlapping of the interlayer orbitals. The calculated results on the interlayer binding energy per unit area, $E_B$ (Table I), also quantitatively support that BP has the strongest interlayer coupling. It has been suggested that the strong interlayer coupling in BP stems from the electronic hybridization of the lone electron-pairs beyond the van der Waals epitaxy[17].

**Table I** Comparison of the interlayer coupling in BP, MoS$_2$ and graphene. Here $\alpha$ is the experimental interlayer force constant, $K_{intra}$ ($K_{inter}$) is the intralayer (interlayer) spring force constant deduced from measured phonon frequencies. $E_B$ is the calculated interlayer binding energy per unit area. Three types of van der Waals corrections to the conventional density functional, Optb88-vdW, DFT-D2, and Optb86-vdW, are employed in the calculations, and they produce consistent results.

| | $\alpha$ ($10^{18}$ N/m$^3$) | $K_{intra}$ / $K_{inter}$ | $E_B$ (meV/Å$^2$) | | |
|---|---|---|---|---|---|
| | | | Optb88-vdW | DFT-D2 | Optb86-vdW |
| **black P** | 127[a] | 26 | 31 | 23 | 35 |
| **MoS$_2$** | 29[b] | 100 | 26 | 18 | 27 |
| **graphene** | 12.8[c] | / | 26 | 21 | 26 |

*a: this work; b: ref. 11; c: ref. 10

In summary, we have successfully observed low-frequency interlayer compression modes in BP films over a wide range of thickness from a few layers to tens of layers. The sensitive layer-number dependence of the frequency shift of these modes enables an accurate determination of the number of atomic layers in multilayer BP. A low-frequency branch of the interlayer modes can be well described by a linear-chain model, which has been used to describe few-layer MoS$_2$ and graphene. In addition, we observed a novel ultralow-frequency collective compression mode that scales with layer number N as $1/\sqrt{N}$, which is a clear indication of an unusually strong interlayer coupling. This surprising result, which is unprecedented in similar 2D layered materials, is further supported by first-principles charge calculations and a phonon frequency analysis. The obtained ratio of ~26 for the intra- and inter-layer force constant in BP is almost four times smaller than that in $MoS_2$. This strong interlayer coupling has significant implications for understanding and modeling of the electronic and mechanical properties crucial to applications of BP-based nanodevices.

## Methods

**Sample preparation**. The BP flakes with various thicknesses were obtained by mechanical exfoliation from bulk crystals synthesized under high pressure. The freshly exfoliated flakes were quickly transferred to 300nm thick $SiO_2$ substrate on a Si wafer. The optical microscope and AFM imaging were employed to determine the thickness. The measurements were controlled within 15 minutes as the BP samples are very easy to be oxidized in air. Then the flakes were covered with PMMA protective film for further Raman measurements. For the flakes without AFM imaging, the PMMA film was immediately covered after exfoliation to keep a high-quality flake surface.

**Raman measurements and AFM imaging.** Raman measurements were performed with a Jobin Yvon HR800 single-grating-based micro-Raman system equipped with a volume Bragg grating low-wavnumber suite, a liquid-nitrogen cooled back-illuminated CCD detector and a 633 nm laser (Melles Griot). The laser was focused into a spot of ~5 μm in diameter on sample surface, with a power less than 100 μW. AFM imaging was carried out with a Nanoscope IIIa Dimension 3100 AFM system (Digital Instruments).

**First-principles calculations**. First-principles calculations were carried out with the Vienna Ab initio Simulation Package (VASP)[26,27], which makes use of the projector augmented wave method[28,29]. The computational details are given in the supplementary materials.

## Acknowledgements

This work was supported by the Ministry of Science and Technology of China

(973 projects: 2012CB921701, 2012CB921704 and 2011CBA00112 ) and the NSF of China. Q.M.Z. & K. L was supported by the Fundamental Research Funds for the Central Universities, and the Research Funds of Renmin University of China (10XNI038 & 14XNLQ03). Computational resources have been provided by the PLHPC at RUC. The atomic structures were prepared with the XCRYSDEN program[30].

# Supplementary Information

# Ultralow-frequency collective compression mode and strong interlayer coupling in multilayer black phosphorus

Shan Dong,[1] Anmin Zhang,[1] Kai Liu,[1] Jianting Ji,[1] Y. G. Ye,[2] X. G. Luo[2,6] and X. H. Chen[2,5,6] Xiaoli Ma,[1] Yinghao Jie,[1] Changfeng Chen,[3] Xiaoqun Wang[1,4] and Qingming Zhang[1,4*]

[1]Department of Physics, Beijing Key Laboratory of Opto-electronic Functional Materials & Micro-nano Devices, Renmin University of China, Beijing 100872, P. R. China

[2]Hefei National Laboratory for Physical Sciences at Microscale and Department of Physics, University of Science and Technology of China and Key Laboratory of Strongly-coupled Quantum Matter Physics, Chinese Academy of Sciences, Hefei, Anhui 230026, China

[3]Department of Physics and High Pressure Science and Engineering Center, University of Nevada, Las Vegas, Nevada 89154, USA

[4]Department of Physics and Astronomy, Collaborative Innovation Center of Advanced Microstructures, Shanghai Jiao Tong University, Shanghai 200240, P. R. China

[5]High Magnetic Field Laboratory, Chinese Academy of Sciences, Hefei, Anhui 230031, China

[6]Collaborative Innovation Center of Advanced Microstructures, Nanjing 210093, China

1. Confirmation of layer number assignment by high-frequency $A_g^2$ mode
2. Calculated low-frequency optical modes in few-layer BP
3. Raman spectra of the BP flakes with/without PMMA glue covering
4. Leakage of the infrared $B_{1u}$ mode
5. Details of the first-principles calculations

# 1. Confirmation of layer number assignment by high-frequency $A_g^2$ mode

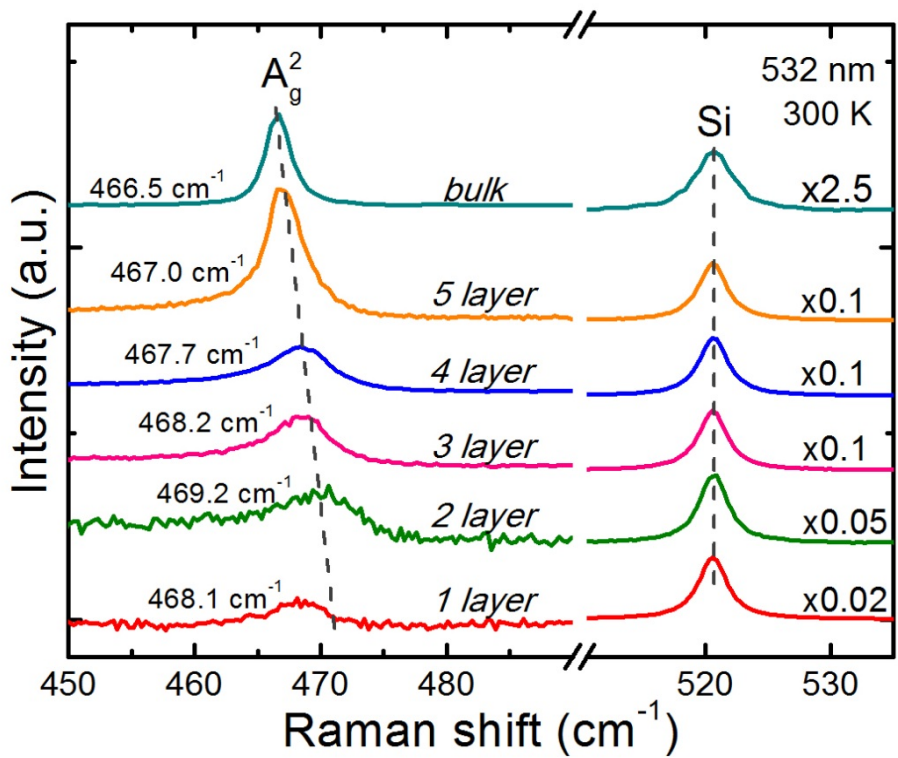

**Fig. S1 High-frequency Raman spectra of few-layer BP.** The evolution of the $A_g^2$ Raman mode of few-layer BP with thickness. The 520.59 cm$^{-1}$ phonon mode from substrate silicon is also shown here for frequency calibration. The laser wavelength is 532 nm (2.33 eV).

To corroborate the layer number assignment based on the low-frequency Raman modes (see the main text), we also performed high-frequency Raman measurements for few-layer BP. The $A_g^2$ modes for the one- to five-layer flakes and bulk BP are shown in Fig. S1. For accuracy, the peak positions are carefully calibrated with the substrate silicon mode located at 520.59 cm$^{-1}$. The systematic frequency shift of the $A_g^2$ mode with the layer number is consistent with the results of our low-frequency measurements and supports our layer-number determination using the low-frequency compression modes. Determination of the layer number using the peak shift of the $A_g^2$ mode has been proposed previously[1], and the anomalous shift of the observed mode towards low frequencies in monolayer BP was attributed to the Davydov splitting of the monolayer vibrational modes, which causes the appearance of new Raman-allowed modes. The observed mode in monolayer BP is actually the $B_{2u}^1$ mode at 468 cm$^{-1}$, rather than the $A_g^2$ mode at slightly higher position[2]. Compared to the much large frequency shifts of the compression modes with layer number (~19 and 32 cm$^{-1}$), the shift of the $A_g^2$ mode is very small (~1.6 cm$^{-1}$). These results demonstrate that the low-frequency compression modes have significant advantages in accurately determining the layer number of multilayer BP.

## 2. Calculated low-frequency optical modes in few-layer BP

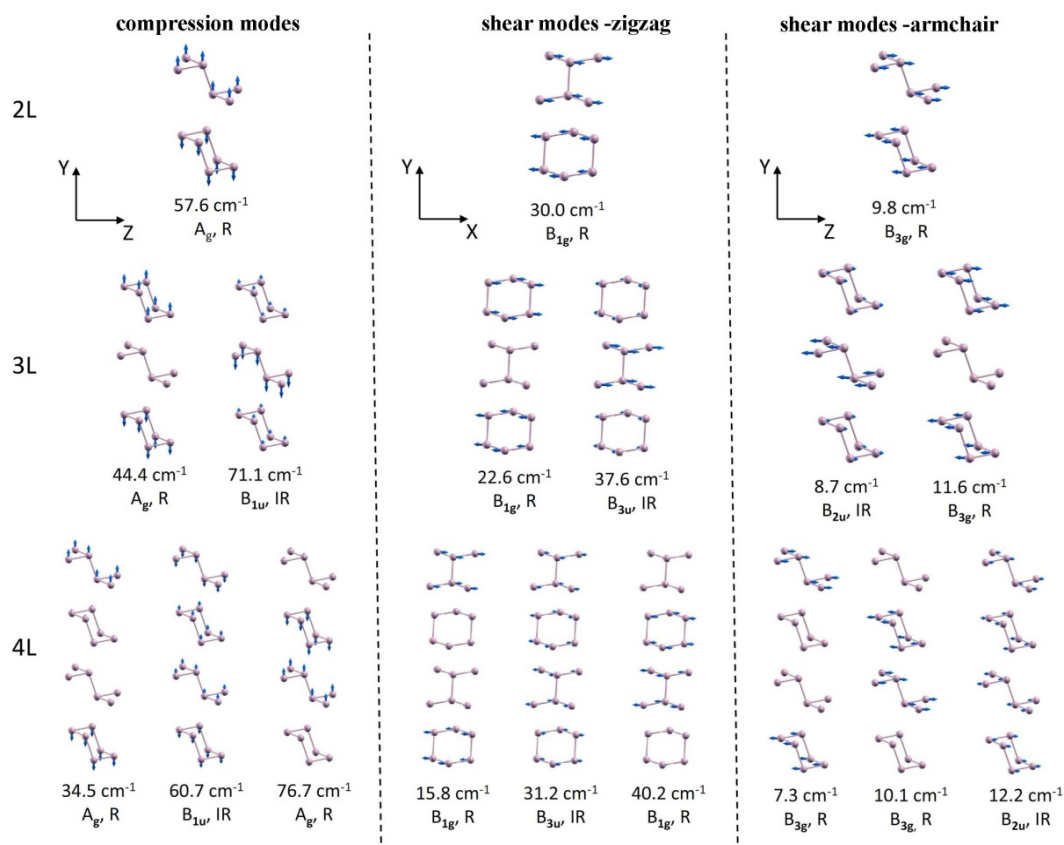

**Fig. S2 Eigen-vectors and frequencies of the optical modes in few-layer BP.** The left, middle and right panels show the vibration patterns of the compression modes, shear modes along the zigzag direction and shear modes along the armchair direction, respectively.

We have calculated the optical modes in few-layer BPs from first principles. The vibration patterns and frequencies of the low-frequency modes are shown in Fig. S2. Only N-1 compression/shear modes are allowed for N-layer BP. In experiments, both the incident and scattered light propagate along the Y axis (see Fig. S2). Under this configuration, the shear modes are experimentally invisible despite that some of them are Raman-active due to the constraints on Raman tensors for the $D_{2h}$ point group.

## 3. Raman spectra of the BP flakes with/without PMMA glue covering

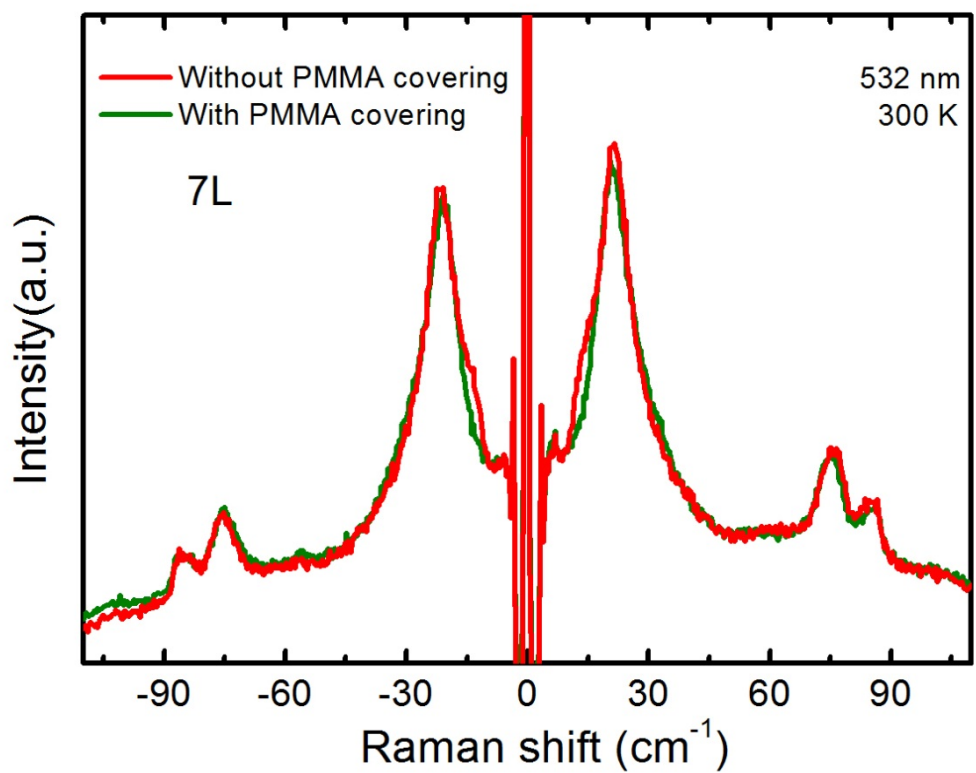

**Fig. S3 Effect of PMMA glue on the low-frequency Raman spectra.** The Stokes and anti-Stokes Raman spectra of the BP flakes with (green) and without (red) protective PMMA glue covering.

Fresh BP flakes are easily degraded in air. Transparent PMMA glue is a good protection for the fresh surface of BP flakes. We have carefully checked the effect of PMMA glue on the low-frequency Raman spectra of a thin (seven-layer) BP flake, as shown in Fig. S3. The almost overlapping spectra indicate that PMMA glue has negligible effect on the low-frequency Raman measurements, in both frequency and intensity.

## 4. Leakage of the infrared $B_{1u}$ mode

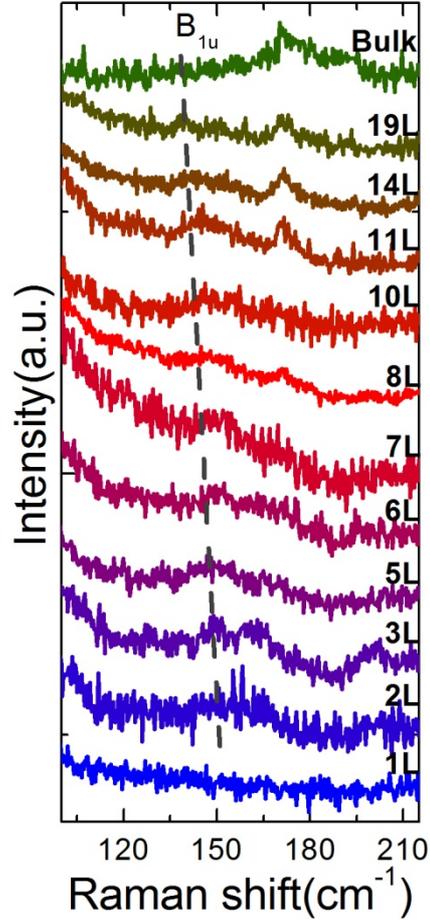

**Fig. S4 The infrared $B_{1u}$ mode in few-layer BP.** The dashed line guides the eyes and approximately traces the evolution of the $B_{1u}$ mode with the layer number.

We observed a weak mode around ~ 139 cm$^{-1}$ in a 19-layer BP flake. It is very close to the infrared $B_{1u}$ mode at 138 cm$^{-1}$ observed in bulk BP[2]. This suggests that a very small mode-leakage may exist in our measurements. It may be related to the defects caused by the high activity of BP. A systematic evolution of the $B_{1u}$ mode with the layer number can also be seen.

## 5. Details of the first-principles calculations

In our first-principles calculations, we adopted the generalized gradient approximation (GGA) of Perdew-Burke-Ernzerh[3] for the exchange-correlation potential. To describe the van der Waals (vdW) interaction in a layered material like BP, which is not included in the conventional density functional theory, we have chosen three different functional forms: the DFT-D2 method[4], the optB86b-vdW functional, and the optB88-vdW functional[5]. We first did the structure optimization of the bulk. Both the cell parameters and internal atomic positions were allowed to relax until the forces were smaller than 0.005 eV/Å. To model the few-layer system, we used a two-dimensional supercell plus a vacuum layer of greater than 10 Å in length. The in-plane lattice constants of the few-layer system were fixed at their equilibrium bulk values and the out-of-plane lattice parameters were fully relaxed. The interlayer binding energy was calculated according to the method described in Ref. 6. The charge difference density was calculated from $\Delta Q = Q(nL) - nQ(L)$, where $Q(nL)$ is the charge density of the n-layer system and $Q(L)$ is the charge density of the isolated monolayer at fixed atomic positions in the n-layer system. The frequencies and displacement patterns of phonon modes were calculated by using the dynamical matrix method.